\documentclass{aa}  

\usepackage{graphicx}
\usepackage{txfonts}
\usepackage{mathrsfs}  
\let\mathcal=\mathscr
\usepackage[colorlinks,allcolors=blue]{hyperref}

\begin{document} 

\title{
Linking star formation thresholds and truncations \\ in the thin and thick disks of the low-mass galaxy UGC~7321
}
\titlerunning{Truncations in the thin and thick disks of UGC~7321}
  \author{S. D\'iaz-Garc\'ia\inst{1,2,3,4}
          \and 
          S. Comer\'on\inst{2,1}
          \and
          S. Courteau\inst{3}
          \and 
          A. E. Watkins\inst{5}
          \and  
          J. H. Knapen\inst{1,2}
          \and  
          J. Rom\'an\inst{1,2,6}
          }
  \institute{Instituto de Astrof\'isica de Canarias, E-38205, La Laguna, Tenerife, Spain  \\
              \email{simondiazgar@gmail.com}
         \and
             Departamento de Astrof\'isica, Universidad de La Laguna, E-38205, La Laguna, Tenerife, Spain
         \and             
             Department of Physics, Engineering Physics \& Astrophysics, Queen's University, Kingston, ON K7L 3N6, Canada
         \and             
             Personal Docente, Consejer\'ia de Educaci\'on, Universidades, Cultura y Deportes del Gobierno de Canarias, E-35002, Las Palmas de Gran Canaria, Spain
         \and
                        Centre for Astrophysics Research, School of Physics, Astronomy 
                        and Mathematics, University of Hertfordshire, Hatfield AL10 9AB, UK
                 \and
                        Kapteyn Astronomical Institute, University of Groningen, PO Box 800, 9700 AV Groningen, The Netherlands
             }
  \date{Received 14 October 2021; accepted 29 August 2022}
\abstract
{
Thin and thick disks are found in most spiral galaxies, yet their formation scenarios remain uncertain. 
Whether thick disks form through slow or fast, internal or environmental, processes is unclear. 
The physical origin of outer truncations in thin and thick disks, observed as a drop in optical and 
near-infrared (NIR) surface brightness profiles, is also a much debated topic. 
These truncations have been linked to star formation (SF) thresholds in Milky-Way-type galaxies, 
but no such connection has been made for their low-mass counterparts or in thick disks. 
Our photometric analysis of the edge-on galaxy UGC~7321 offers a possible breakthrough. 
This well-studied diffuse, isolated, bulgeless, ultra-thin galaxy is thought to be under-evolved both dynamically and in SF. 
It is an ideal target for disentangling internal effects in the formation of thick disks and truncations. 
Our axial light profiles from deep far- and near-ultraviolet ({GALEX}) images, 
tracing recent SF, and optical (DESI $grz$) and NIR (\emph{Spitzer} 3.6 $\mu$m) images, tracing old stellar populations, 
enable a detailed identification of an outer truncation in all probed wavelengths in both the thin and thick disks.
After deprojecting to a face-on view, a sharp truncation signature is found 
at a stellar density of $1.5 \pm 0.5 \, \mathcal{M}_{\odot} \, {\rm pc}^{-2}$, 
in agreement with theoretical expectations of gas density SF thresholds. 
The redder colours beyond the truncation radius are indicative of stellar migration towards the outer regions. 
We thus show that thick disks and truncations can form via internal mechanisms alone, given the pristine nature of UGC~7321. 
We report the discovery of a truncation at and above the mid-plane of 
a diffuse galaxy that is linked to a SF threshold; 
this poses a constraint on physically motivated disk size measurements among low-mass galaxies. 
}

\keywords{galaxies: individual UGC~7321 - galaxies: structure - galaxies: star formation}

\maketitle

\section{Introduction}\label{introduction}

The presence of sharp drops in the outer parts of the surface brightness (SB) profiles of some edge-on galaxies 
has been known for decades \citep[][]{1970ApJ...160..811F, 1979A&AS...38...15V}. 
The formation of these so-called truncations has been studied extensively in galaxies of different inclinations 
\citep[e.g.][]{2006A&A...454..759P,2008MNRAS.386.1821F,2012ApJ...758...41R,2012ApJ...759...98C,2016MNRAS.456.1359F} 
and redshifts \citep[$z$; e.g.][]{2004A&A...427L..17P}. 

Truncations have been linked to a critical gas surface density for star formation \citep[SF;][]{2001ApJ...555..301M}. 
Examples include the Milky Way (MW)-like edge-on galaxies NGC~4565 and NGC~5907, 
whose large angular sizes allow for a photometric analysis with high spatial resolution. 
\citet[][]{2019MNRAS.483..664M} identified truncations in their disks, at heights of up to $\sim 3$~kpc, 
from near-ultraviolet (NUV; tracer of recent SF), optical (stacked $gri$), and near-infrared 
(NIR; 3.6 $\mu$m, tracer of old stellar populations) wavelengths. 
Truncations in these two galaxies lie at a stellar surface density ($\Sigma_{\star}$) 
of $\sim 1-2 \, \mathcal{M}_{\odot} \, {\rm pc}^{-2}$ \citep[][]{2019MNRAS.483..664M}, 
which is consistent with the critical gas surface density ($\sim 3-10 \, \mathcal{M}_{\odot} \, {\rm pc}^{-2}$) 
beyond which gas can no longer be transformed into stars \citep[][]{2004ApJ...609..667S}. 
In this paper we extend this work into the realm of low-mass galaxies with an analysis of the 
truncated diffuse edge-on galaxy UGC~7321 
\citep[maximum circular velocity $V_{\rm c}=108 \, {\rm km} \, {\rm s}^{-1}$;][]{2005ApJS..160..149S}.

\citet[][]{2020A&A...633L...3C} and \citet[][]{2020MNRAS.493...87T} proposed the radius corresponding to 
the isomass contour at $1 \, \mathcal{M}_{\odot} \, {\rm pc}^{-2}$ 
as a physically motivated disk size measurement ($R_{1}$) for face-on and moderately inclined galaxies. 
This is supported by the aforementioned $\Sigma_{\star}$ values at the truncation of MW-type galaxies. 
The use of $R_{1}$ yields a narrower stellar mass–size relation than the half-light radius 
\citep[see also][]{2020MNRAS.495...78S,2021MNRAS.505.3135A,2022A&A...660A..69W}. 
Similar conclusions were reached by \citet[][]{2015ApJS..219....3M} using the 25.5 mag arcsec$^{-2}$ isophotal radius 
from 3.6 $\mu$m images from the \emph{Spitzer} Survey of Stellar Structure in Galaxies \citep[S$^4$G;][]{2010PASP..122.1397S}. 
Below, we study $\Sigma_{\star}$ at the edge of a low-mass low-SB galaxy to further constrain its size.

Thick disks --- the faint, large scale-height counterpart of thin discs --- are frequently 
identified in edge-on galaxies and 
can also present truncations \citep[e.g.][]{2011ApJ...741...28C}. 
They are dominant in low-mass galaxies \citep[][]{2006AJ....131..226Y}. 
Their formation, whether fast or slow, internal or driven by the environment, is also a point of contention. 
Thick disks could be the fossil record of a primordial turbulent disk \citep[e.g.][]{2006ApJ...650..644E} 
built through an intense SF episode \citep[][]{2014A&A...571A..58C} 
or quickly formed at high $z$ through wet mergers \citep[e.g.][]{2004ApJ...612..894B}. 
They could also gradually arise from stars stripped from, and dynamically heated by, 
infalling satellites \citep[][]{2003ApJ...597...21A} or via internal 
mechanisms such as radial migration \citep[][]{2009MNRAS.399.1145S} and 
dynamical heating by giant molecular clouds \citep[][]{1985ApJ...290...75V}. 
Thick disk formation can result from the simultaneous interplay of internal and external processes 
\citep[][]{2015ApJ...804L...9M,2019A&A...623A..19P}, though we now focus on the former 
by studying a galaxy (UGC~7321) without a strong sign of environmental activity.

UGC~7321 is catalogued as an isolated galaxy by \citet[][]{2005A&A...436..443V} and \citet[][]{2017A&A...603A..18H}. 
NGC$\,$4204 appears at an offset of $\approx 2^{\circ}$ southwards, 
and NGC$\,$4455 lies at an even larger angular distance, 
though both galaxies are a factor of $\sim 2$ closer to us than 
UGC~7321 \citep[e.g.][]{2006Ap.....49..450K,2011A&A...532A.104N,2017A&A...603A..18H}. 
The isolation of UGC~7321 is further confirmed from the low values of the projected surface density 
to the third nearest neighbour ($\Sigma_{3}^{\rm A}=0.7$) and its Dahari parameter ($Q=-5.1$) \citep[][]{1984AJ.....89..966D} 
in a velocity interval of $\pm \, 500 \, {\rm km \, s}^{-1}$, following \citet[][]{2014MNRAS.441.1992L}. 
Inspection of optical images from the Dark Energy Spectroscopic Instrument (DESI) 
Legacy Imaging Surveys \citep[][]{2019AJ....157..168D} 
does not reveal the presence of any low-SB satellite within an area encompassing a factor of several times the galaxy radius, 
down to depths of $\mu_{g}$(AB)$=28.9$ mag arcsec$^{-2}$ and $\mu_{r}$(AB)$=28.2$ mag arcsec$^{-2}$ 
\citep[$3\sigma$, 10\arcsec x10\arcsec boxes;][]{2020A&A...644A..42R}. 
Hence, the main photometric and kinematic properties of UGC~7321 are likely not affected or caused by interactions. 
The presence of extremely faint dwarfs cannot be discarded, as their detection might demand even deeper imaging 
\citep[see e.g.][]{2017A&A...603A..18H}.

UGC~7321 is also a bulgeless galaxy with a very cold disk \citep[][]{1999AJ....118.2751M}, 
hinting at a quiescent merger history \citep[][]{2019A&A...628A..58S}. 
However, its warped H{\sc\,i} disk is indicative of a possible encounter ($>1.6$ Gyr ago) 
with a neighbour, followed by disk cooling \citep[][]{2003AJ....125.2455U}, 
or angular momentum misalignments between the disk and the dark matter halo \citep[][]{1999ApJ...513L.107D}. 
Such a warp could indeed be of intrinsic origin \citep[see further discussion in][]{2017ASSL..434..209B}. 
UGC 7321's ultra-thin disk \citep[][]{2017MNRAS.465.3784B} suggests a higher 
spin parameter than those of low-SB and regular disk galaxies \citep[][]{2019MNRAS.488..547J} 
or a dark matter dominance \citep[][]{2013MNRAS.431..582B}. 
Also, ultra-thin galaxies have been found in specific cosmic web environments with a very low density, 
as they are less connected with filaments \citep[][]{2017MNRAS.465.3784B}. 
Altogether, UGC~7321 appears to be extremely isolated and 
under-evolved in terms of dynamics and SF \citep[e.g.][]{1999AJ....118.2751M}.

UGC~7321 has a thick disk. Early studies of its vertical density distribution revealed that a 
single sech$^{2/n}$ fit cannot reproduce optical observations; a second component is needed \citep[][]{2000AJ....120.1764M}. 
The model of \citet[][]{2019A&A...628A..58S} -- including a gravitationally coupled stellar disk and a H{\sc\,i} disk 
in the potential of a dark matter halo \citep[see also][]{2010NewA...15...89B} -- suggests that 
$n$ cannot be trusted as a robust parameter, as it varies with radius and fitting range, and thus a double-disk fit may not be necessary. 
This model is, on the other hand, based on the relatively shallow optical surface photometry 
from \citet[][]{1999AJ....118.2751M}. Even so, a thick disk was already detected in the Matthews et al. $B$-$R$ colour maps. 
The deeper 3.6~$\mu$m imaging from the S$^4$G and distance-independent 
photometric decomposition models of \citet[][]{2018A&A...610A...5C} confirm that a thick disk is definitely needed to 
fit the vertical SB profiles (see their Appendix B). 
Comer\'on et al. assumed fitting functions for two stellar disks and 
one gaseous isothermal coupled disk in equilibrium, and their fit is used in this article.

In addition, a truncation in both the thin and thick disks is detected in the 3.6~$\mu$m SB axial\footnote{The 
axial direction is the mid-plane projection of a vector pointing away from the galaxy centre in the sky plane.} 
profiles by \citet[][]{2018A&A...610A...5C}. 
Here, we revisit this truncation in NUV, far-ultraviolet (FUV), $grz$, and 3.6 $\mu$m NIR images of UGC~7321 
in order to study its connection to SF thresholds. 
Such a connection found in thick disks would constrain formation scenarios, given their expected old age. 
For comparison with MW-type galaxies, the SB profiles of NGC~4565 
\citep[$V_{\rm c}=250 \, {\rm km} \, {\rm s}^{-1}$;][]{2005ApJS..160..149S} 
are also revisited; further details are available in \citet[][]{2019MNRAS.483..664M} and Mart\'inez-Lombilla et al. (in prep.). 
We adopt redshift-independent distances of $22.28 \pm 3.34$ and $13.43 \pm 2.01$ Mpc for UGC~7321 and 
NGC~4565, respectively \citep[][]{2016AJ....152...50T}, assuming a $15\%$ uncertainty \citep[][]{2015ApJS..219....3M}.

\section{Ultraviolet, optical, and near-infrared imaging}\label{data}
\begin{figure}
\centering
\includegraphics[width=0.49\textwidth]{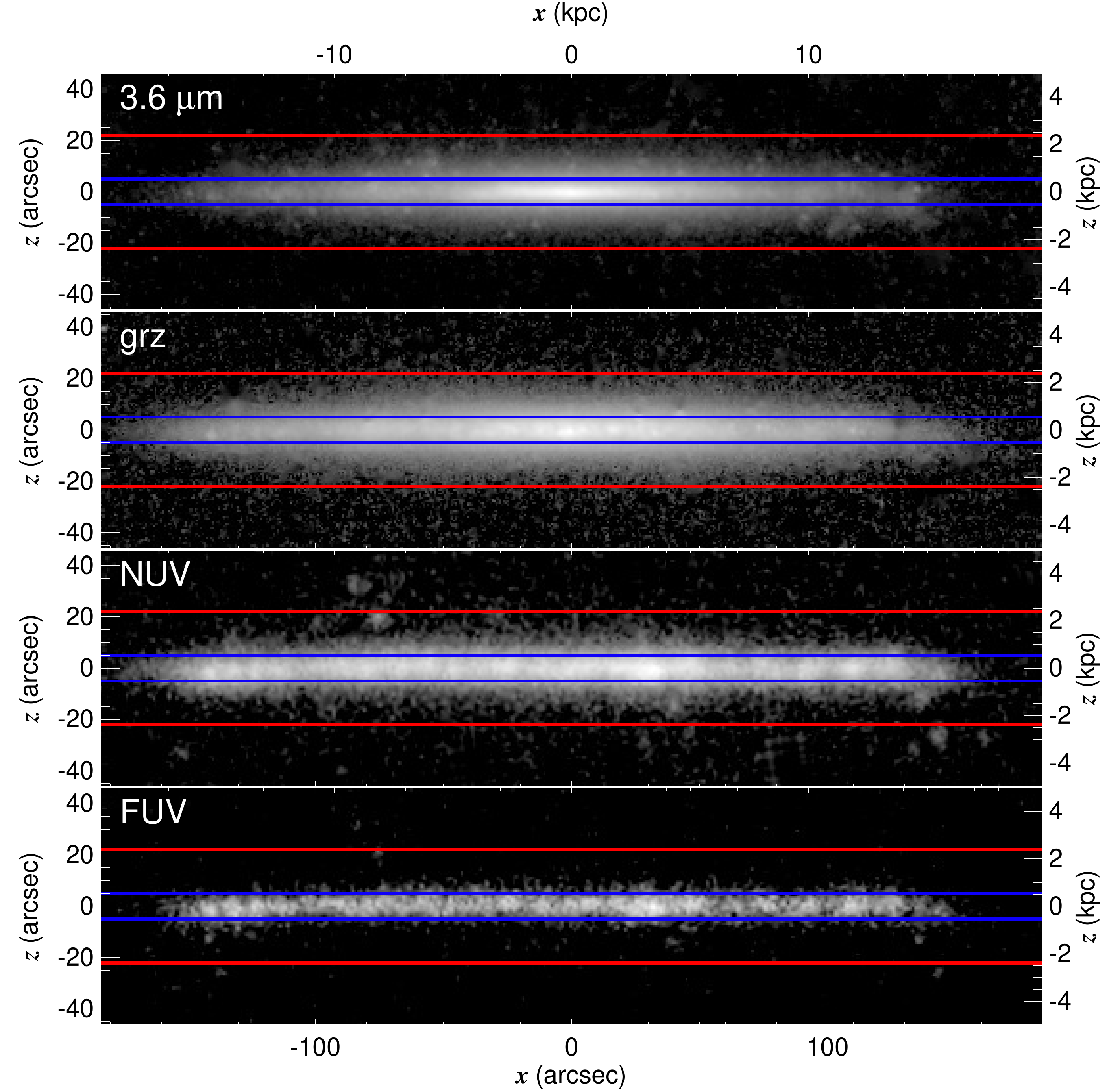}
\caption{
From top to bottom, rotated sky-subtracted images of the low-mass ultra-thin galaxy UGC~7321 in 
3.6 $\mu$m, $grz$ (stacked), NUV, and FUV. The red lines indicate the height at which SB equals 26 mag arcsec$^{-2}$ in 3.6 $\mu$m, 
while the blue lines show the height (5\arcmin) above which $90\%$ of the light comes from the thick disk. 
These values are averaged from the vertical luminosity cuts of \citet[][]{2018A&A...610A...5C}.
}
\label{images_UGC07321_NGC4565_all}
\end{figure}
To trace old stellar populations \citep[][]{2014RvMP...86...47C,2015MNRAS.452.3209R}, 
we used 3.6 $\mu$m images from the S$^{4}$G obtained with the Infrared Array Camera \citep[][]{2004ApJS..154...10F} 
on board the \emph{Spitzer} Space Telescope \citep{2004ApJS..154....1W}, with an exposure time of 240~s per galaxy. 
We also used stacked \text{DESI} $g$, $r$, and $z$ images, with nominal exposure times of 166, 134, and 200 s, 
respectively \citep[][]{2019AJ....157..168D}.

We probed recent SF with the Galaxy Evolution Explorer (GALEX) ultraviolet (UV) 
images from the catalogue of \citet[][]{2018ApJS..234...18B}. 
Specifically, we used NUV ($\lambda_{\rm eff}=2267\,\AA$) and FUV ($\lambda_{\rm eff}=1516\,\AA$) images with long exposure times: 
for UGC~7321, 1683 and 2822\,s in FUV and NUV, respectively; for NGC~4565, 1693\,s for both. 
The FUV emission traces SF of several tens to $100$~Myr, while the NUV traces 
$\lesssim 300$~Myr populations \citep[][]{1998ARA&A..36..189K}. 

All images used the masks created by \citet[][]{2018A&A...610A...5C}, and 
fluxes in masked regions were interpolated following \citet[][]{2015ApJS..219....4S}. 
The sky levels were measured within the same $30\arcsec$x$30\arcsec$ boxes as in \citet[][]{2015ApJS..219....4S}, 
which are located far from the galaxy. 
The median of the median values of each box was then subtracted from the images.

\section{Axial surface-brightness profiles}\label{surface_cuts}
\begin{figure}
\centering
\includegraphics[width=0.4175\textwidth]{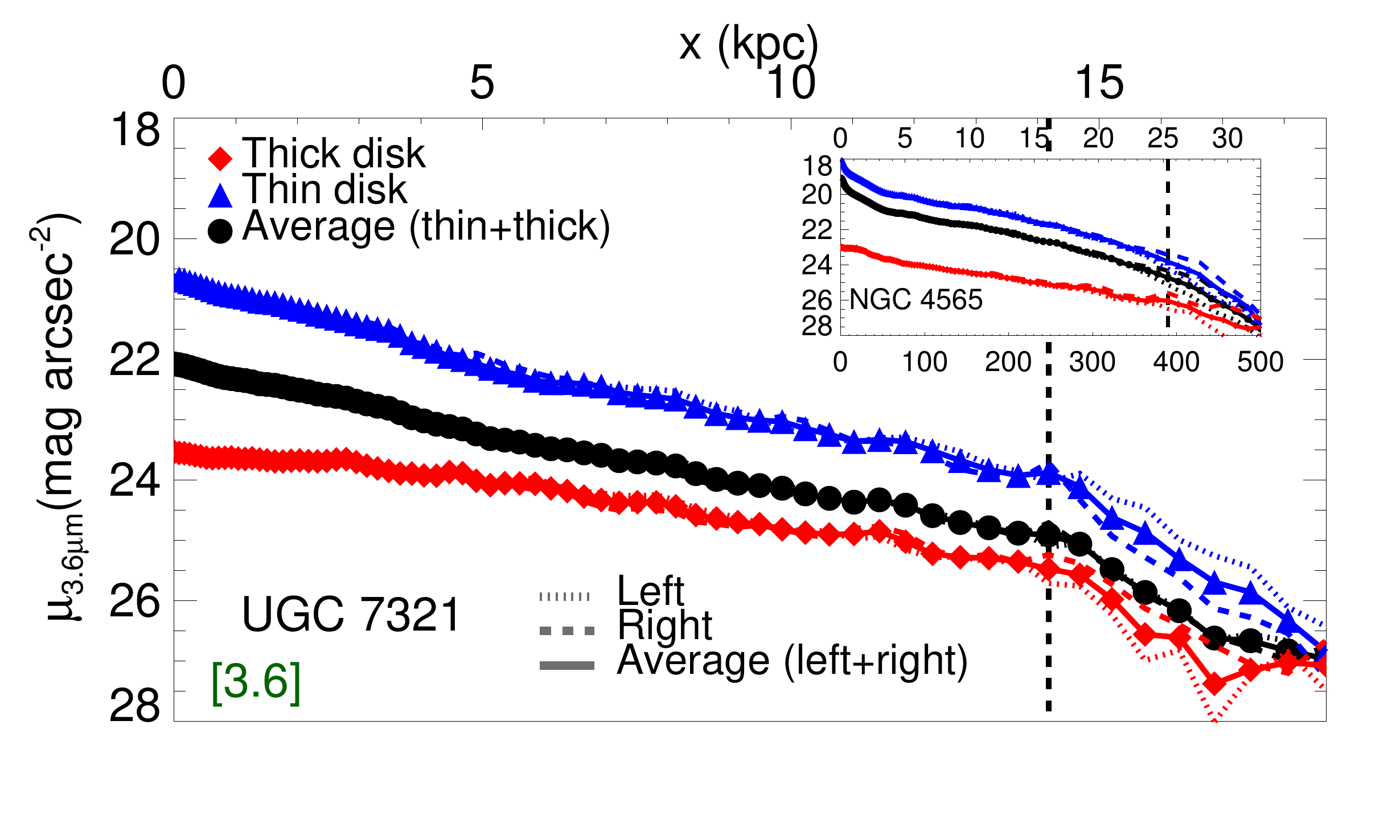}\\[-8.1ex]
\includegraphics[width=0.4175\textwidth]{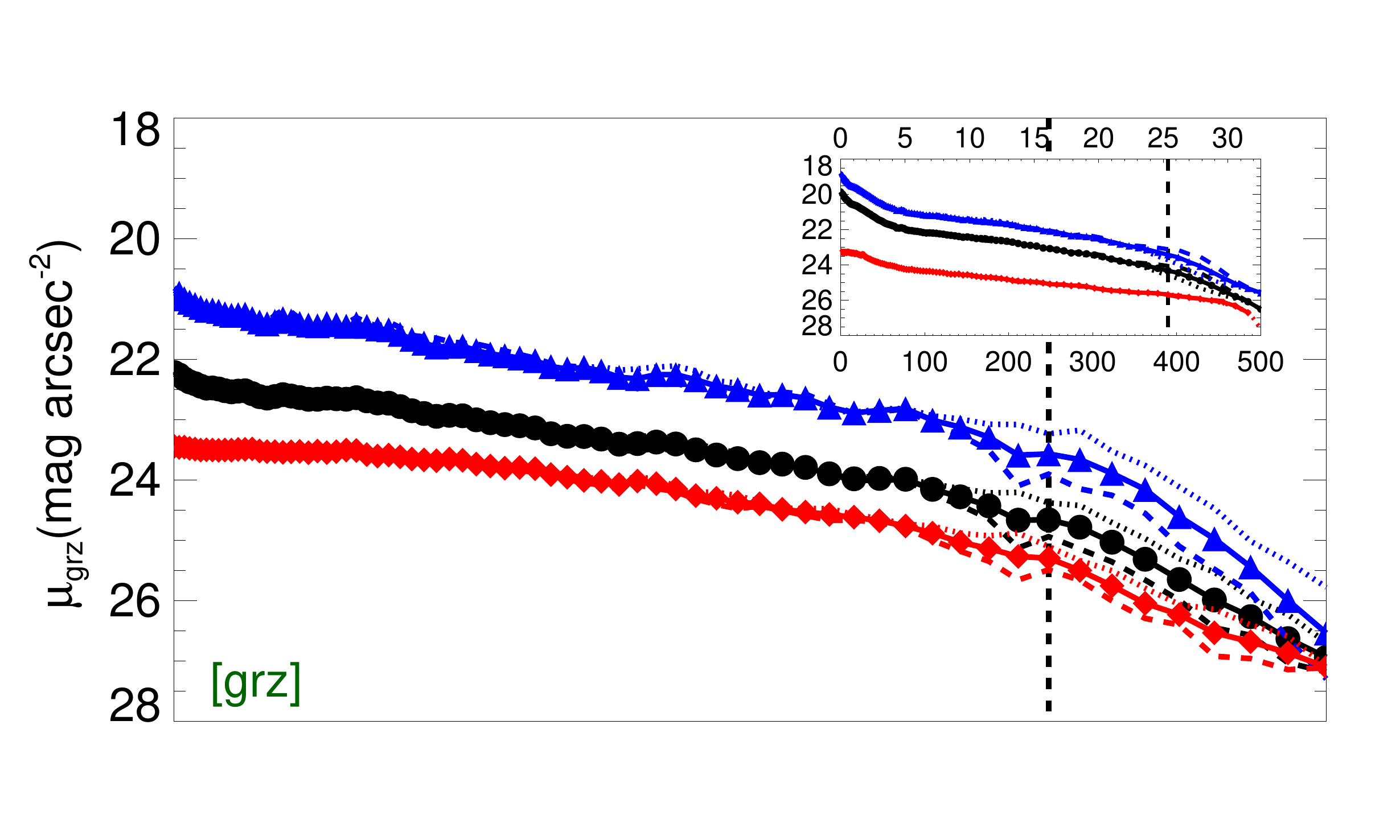}\\[-8.1ex]
\includegraphics[width=0.4175\textwidth]{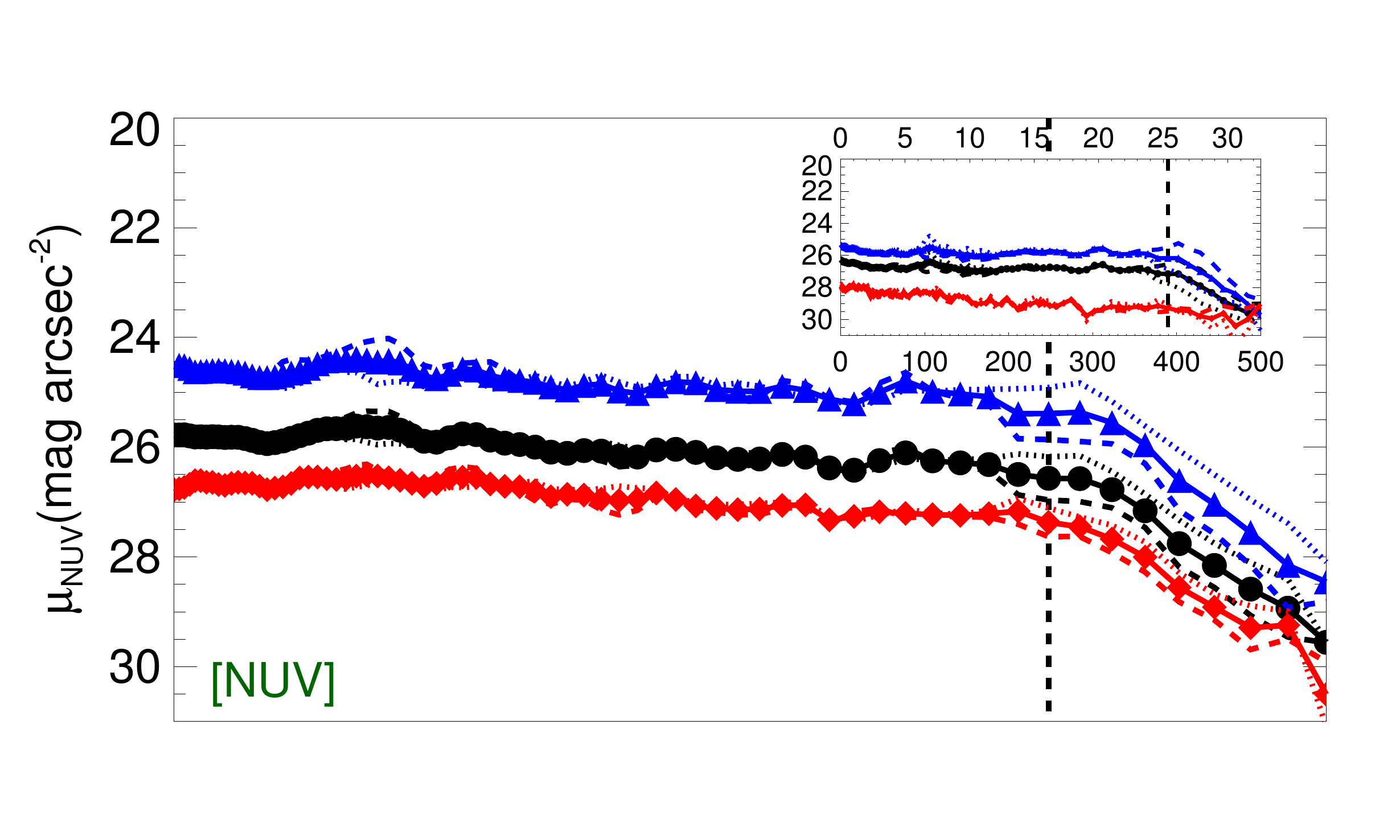}\\[-8.1ex]
\includegraphics[width=0.4175\textwidth]{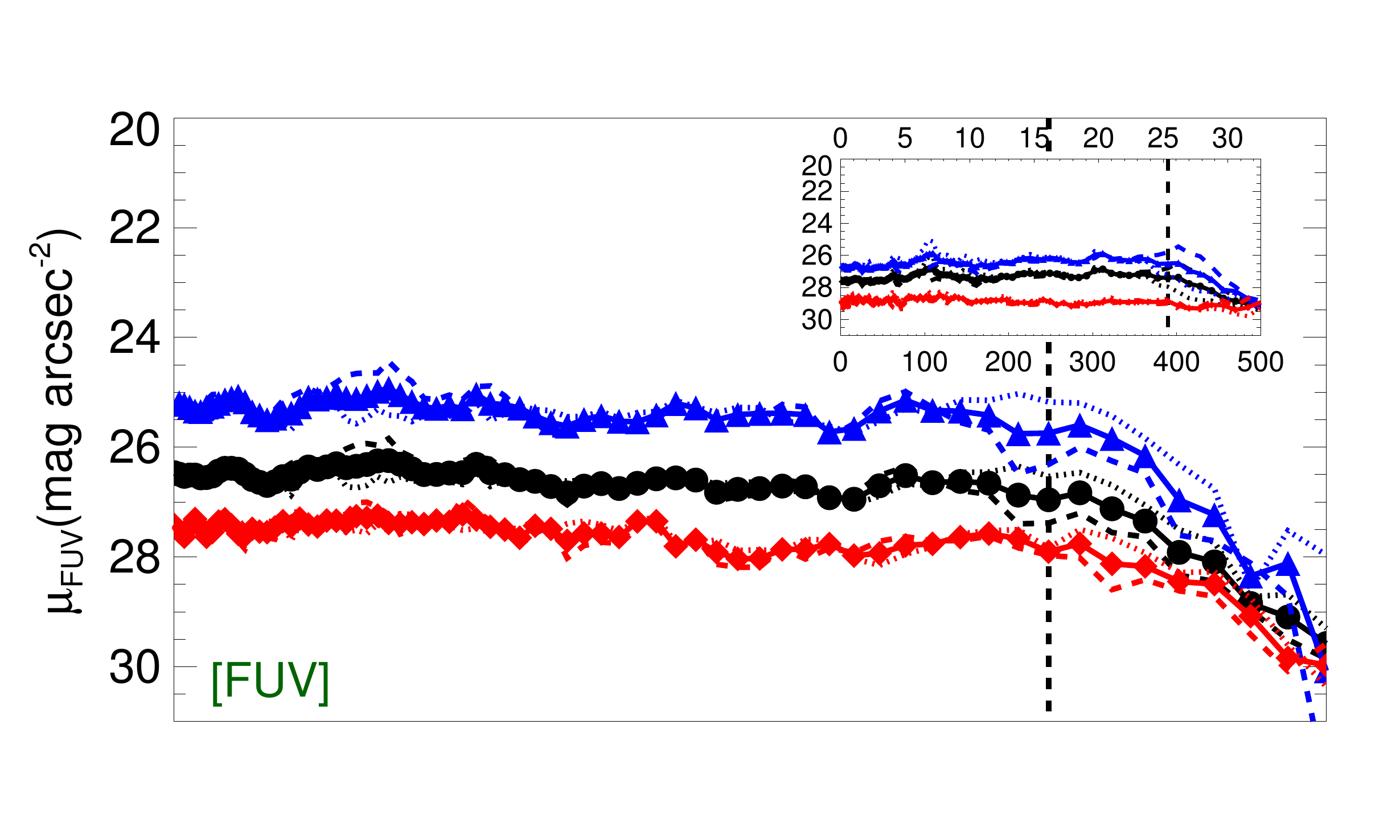}\\[-8.1ex]
\includegraphics[width=0.4175\textwidth]{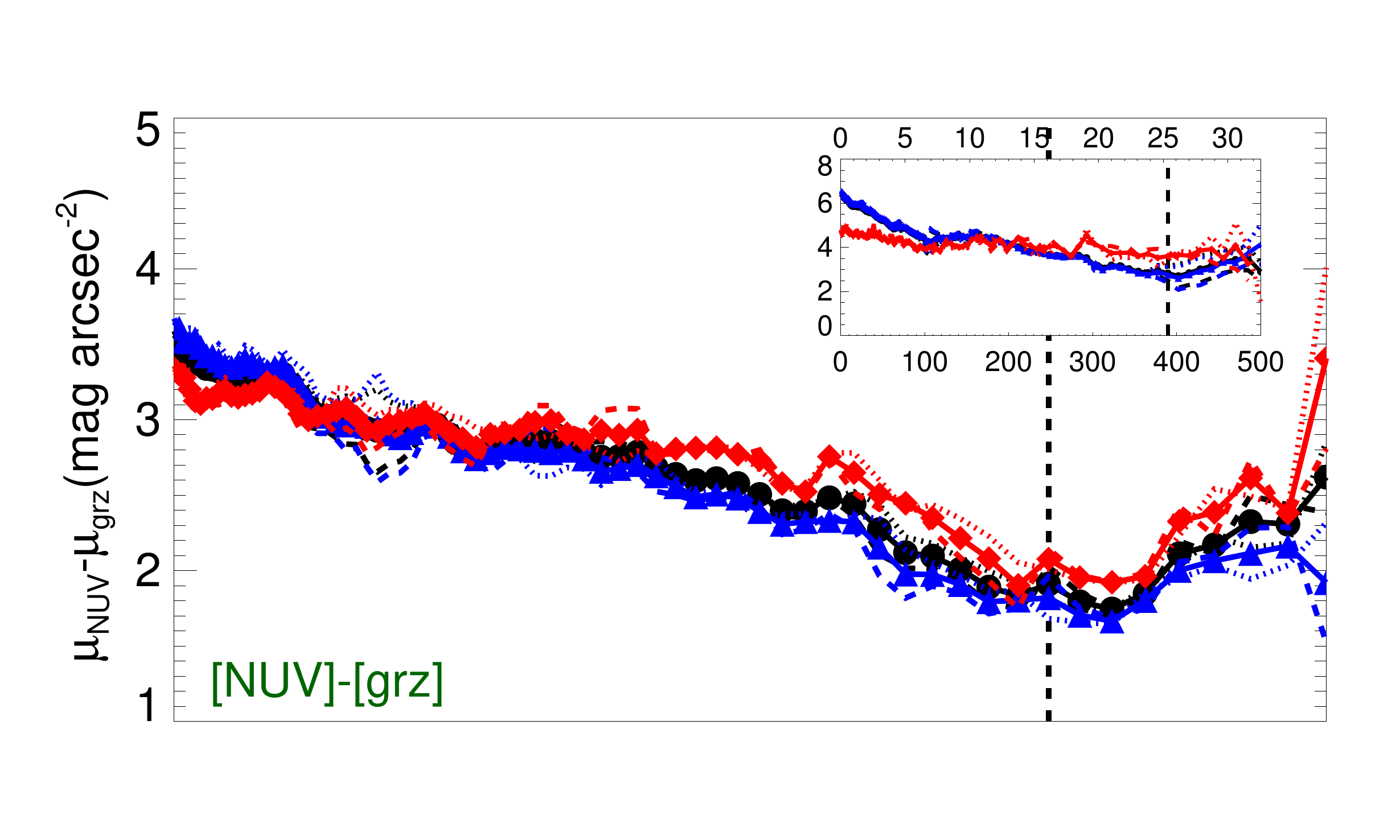}\\[-8.1ex]
\includegraphics[width=0.4175\textwidth]{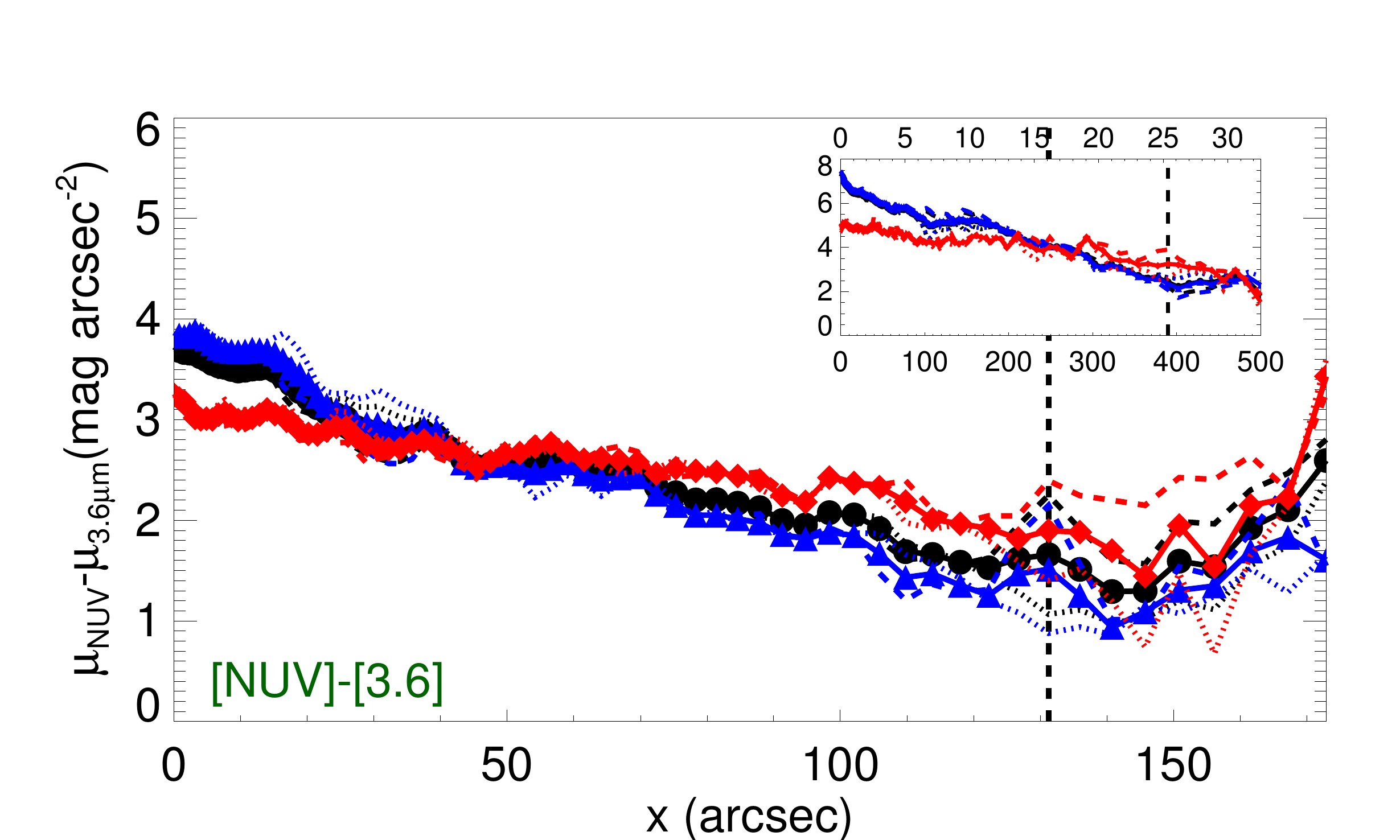}\\
\caption{
Axial SB profiles of UGC~7321 at 3.6 $\mu$m (\textit{first panel}), 
$grz$ combined mean (\textit{second panel}), NUV (\emph{third panel}), and FUV (\emph{fourth panel}), 
along with the NUV-[$grz$] (\emph{fifth panel}) and the NUV-[3.6] (\emph{sixth panel}) colours. 
The radial profiles are shown for the thin disk (in blue), the thick disk (in red), and the entire disk (black). 
For consistency with \citet[][]{2018A&A...610A...5C}, the analysis is limited to their outermost fitted point. 
The inset plots display the same radial profiles for the MW-type NGC~4565 \citep[see also][]{2019MNRAS.483..664M}. 
The vertical dashed lines indicate the truncation loci in the 3.6 $\mu$m images.
}
\label{plot_thin_thick_S4G}
\end{figure}
The GALEX, DESI, and S$^{4}$G images were aligned with the Interactive Data Language (IDL) package \texttt{hastrom}, 
which builds on the \texttt{poly\_2d} function to perform polynomial warping of images. 
The 3.6 $\mu$m images were used as reference to match the astrometry. 
We then obtained axial SB profiles from the 3.6~$\mu$m, $grz$, FUV, and NUV images (Fig.~\ref{images_UGC07321_NGC4565_all}) 
by folding the images with respect to the mid-plane and averaging the flux between heights $z=0$ and $z=z_{\rm u}$, 
where $z_{\rm u}$ is the height at which $\mu_{3.6 \mu \rm m}$(AB) = 26 mag arcsec$^{-2}$ on average. 
The latter was measured from vertical SB profiles presented in \citet[][]{2018A&A...610A...5C}. 
The resulting SB values are sensitive to the selection of $z_{\rm u}$ --- which was 
chosen to maximise the disk light and minimise the background noise --- and are only used for the detection of the truncations. 
The axial profiles were then converted to face-on radial profiles (Sect.~\ref{inclination_correction}). 

Following \citet[][]{2006A&A...454..759P}, we used a logarithmic binning in the axial direction: 
each range is 1.03 times wider than the previous, where the first data point is located at the S$^{4}$G pixel size ($0.75\arcsec$). 
Surface brightness profiles were folded with respect to the galaxy minor axis and averaged 
in the axial direction (asymmetries are discussed in Sect.~\ref{7321_truncation}). 
Vertical 3.6~$\mu$m SB profiles were decomposed by \citet[][]{2018A&A...610A...5C} into thin and thick disks. 
As in their work, we hereafter consider the thin (thick) disk as the region below (above) 
the height at which $90\%$ of the light comes from the thick disc (blue lines in Fig. \ref{images_UGC07321_NGC4565_all}). 

A characterisation of the point spread function (PSF) may be required to reveal the faintest 
stellar structures \citep[][]{2017MNRAS.470..427P,2019A&A...629A..12M,2020MNRAS.491.5317I,2020A&A...644A..42R}. 
We did not consider PSF modeling in NIR and optical wavelengths, however, as we are not probing the 
dimmest regions of thick disks ($> 26$ mag arcsec$^{-2}$) in vertical SB profiles, 
where PSF effects become dominant \citep[see e.g.][]{2018A&A...610A...5C,2019MNRAS.483..664M}.

In order to assess whether the high-altitude UV emission in UGC~7321 can be caused by scattered light, 
we estimated the line PSF (LSF) following \citet[][]{2017ApJ...847...14E} \citep[see also][]{2013A&A...556A..54V}. 
We used the {GALEX} NUV and FUV PSF, extended with a power law using the parametrisation by \citet[][]{2016ApJ...833...58H}. 
We compared the LSFs to vertically integrated SB profiles calculated from the inner $30\arcsec$ and 
confirmed that the LSF is much narrower than the UV disk of UGC~7321. 
We then convolved the LSF with a sech$^2$ disk with an exponential scale height of $1.5\arcsec$ \citep{2018A&A...610A...5C} 
and found negligible differences relative to the LSF. We finally verified that the differences between a radial cut of the PSF 
and the LSF arise in the wings. The wings start affecting the LSF 
at $\sim 1/100$ of the peak value, which is below our detection threshold. 
We thus conclude that the scattered UV light from PSF wings is limited to fainter SB levels than those probed in this work.

The outer truncations and their radii were identified using the break-finding algorithm of \citet[][]{2019A&A...625A..36W}. 
Their method looks for significant changes in the mean of the local slope of the SB profile --- obtained 
following \citet[][]{2006A&A...454..759P} --- using a cumulative sum (CS) of the difference from the mean. 
The location of the truncation corresponds to the maximum of the CS. 
The significance of the truncation is tested by bootstrapping the SB profile $10^{5}$ times 
so that it is randomly reordered in the axial direction. The break strength --- measured as max(CS)-min(CS) --- in the reordered 
profiles must also fall below that of the real profile in $>95\%$ of cases.

\subsection{The NIR, optical, and UV truncations of UGC~7321}\label{7321_truncation}

The SB profiles of UGC~7321 have the same outer truncation radius of $131.3 \arcsec$ or $14.2 \pm 2.1 \, {\rm kpc}$ 
(Fig.~\ref{plot_thin_thick_S4G}) at all probed wavelengths (NUV, FUV, stacked $grz$, 3.6$~\mu$m). 
Likewise, we confirmed that truncations hold, at the same radii, 
in 3.4 $\mu$m images from the Wide-field Infrared Survey Explorer \citep[WISE;][]{2011ApJ...735..112J} 
(Prof. T. Jarrett; private communication). 
We verified that the multi-$\lambda$ truncation is not due to a morphological asymmetry: 
it holds if, instead of symmetrising the axial profiles, 
we study separately the left and right sides along the major axis, as displayed in Fig.~\ref{images_UGC07321_NGC4565_all}. 
This is not the case for NGC~4565 \citep[][]{2019MNRAS.483..664M,2020ApJ...897..108G}, 
which, unlike UGC~7321, is known to be interacting \citep[e.g.][]{2012ApJ...760...37Z}.

The truncation in UGC~7321 appears at the same radius in both the thin and thick disks. 
(Since the latter are defined from 3.6~$\mu$m SB decomposition models alone, 
the claim for the existence of a truncated thick disk with UV emission sources cannot be formally made; 
see Sect.~\ref{thick_disk_internal}.) Dust obscuration in the mid-plane is substantial in the UV (and milder in the NIR), 
but correcting for the dust is non-trivial and beyond the scope of this paper. 
As truncations were also identified above the mid-plane at all probed wavelengths, where there is less dust, 
they cannot be artificially created by dust obscuration 
\citep[for a related discussion on the case of NGC~4565, see][]{2019MNRAS.483..664M}. 
Conversely, the dust distribution in galaxies with $V_{\rm c}<120 \, {\rm km} \, {\rm s}^{-1}$ is thought to have 
a larger scale-height and to be more diffuse than in their massive counterparts \citep[][]{2004ApJ...608..189D}. 
Nonetheless, it is known that UGC~7321 is not optically thick and that its internal extinction is low \citep[][]{1999AJ....118.2751M}. 
Also, the observed central UV SB is > 8 mag dimmer than expected from an extrapolation of 
the observed SB profile slope beyond the truncation; such a difference cannot be solely ascribed to dust.

NUV-[3.6] and NUV-[$grz$] colours trace the specific SF rate \citep[e.g.][]{2018ApJS..234...18B}, 
or the ratio of the SF rate to stellar mass surface densities, 
which in turn is related to the star formation efficiency \citep[SFE; e.g.][]{2011MNRAS.415...61S}. 
Colours become redder beyond the truncation of UGC~7321 
(bottom two panels of Fig.~\ref{plot_thin_thick_S4G}; see Sect.~\ref{discussion_section} for a follow-up discussion).

\subsection{Deprojection of the stellar surface density profiles}\label{inclination_correction}

The SB of a highly inclined galaxy is enhanced due to line--of--sight integration of stars. 
Correcting for this effect is necessary for the photometric analysis of 
galaxies~\citep[e.g.][]{1994ApJ...432..114B,2020MNRAS.493...87T,2021ApJ...912...41S}. 
Indeed, to expand the connection between SF thresholds and galaxy edges in low-mass galaxies, 
the stellar surface density at the truncation radius must be computed from deprojected NIR SB profiles. 

\begin{figure}
\centering
\includegraphics[width=0.49\textwidth]{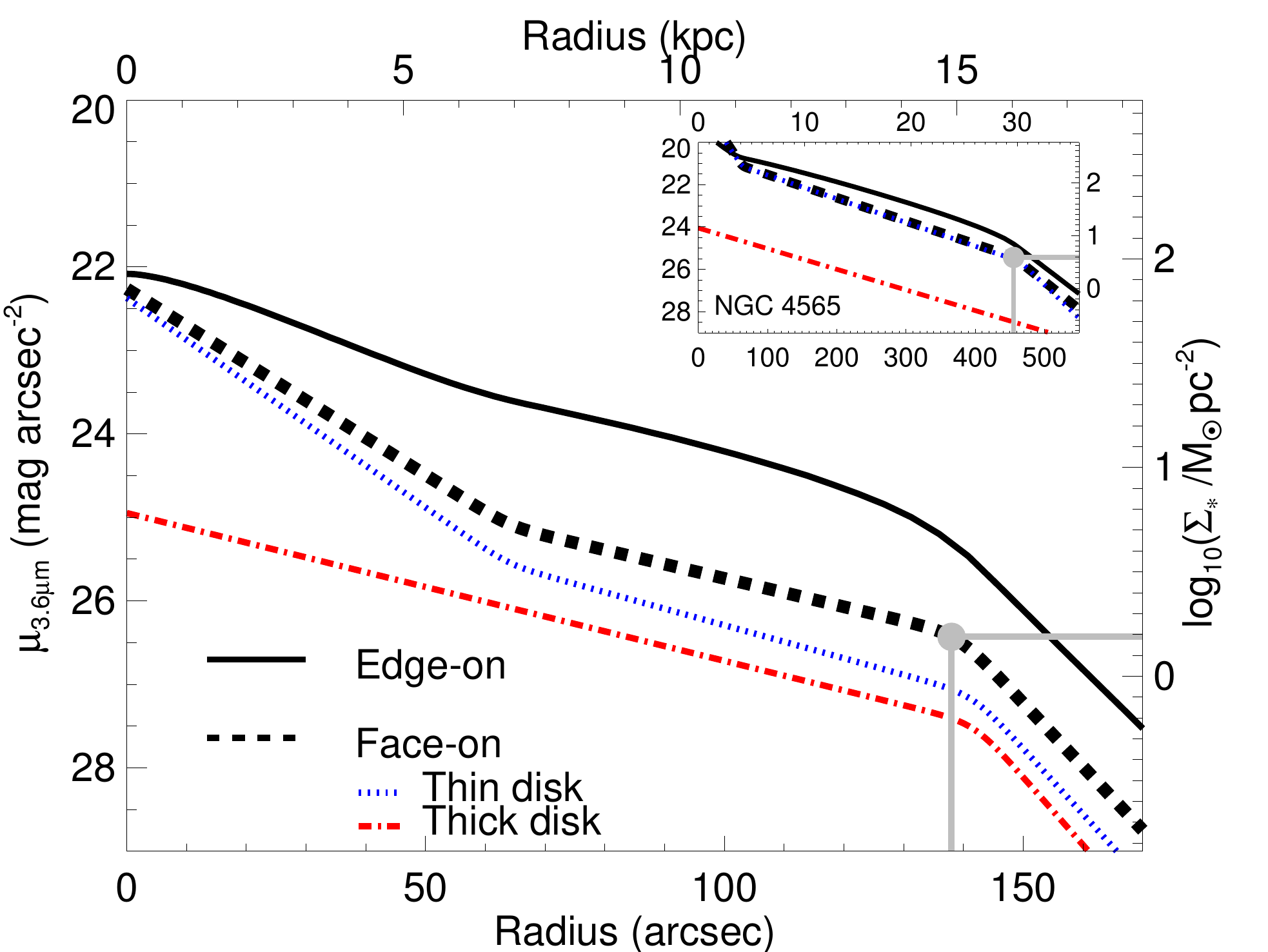}
\caption{
1D model of the 3.6 $\mu$m axial SB profiles of UGC~7321 in edge-on (solid black line) and face-on (dashed black line) views. 
Blue (dotted) and red (dash-dotted) lines correspond to the deprojected radial profiles for the thin and thick disks, respectively. 
The right $y$ axis displays the conversion to surface stellar densities. 
Vertical and horizontal grey lines highlight the $y$ and $x$ intercepts (in the galactic plane) at the truncation. 
The inset shows the same profiles for NGC~4565.
}
\label{plot_thin_thick_deprojections}
\end{figure}

The conversion of axial 3.6 $\mu$m SB profiles ($\mu_{3.6\mu \rm m}$) to 
face-on radial profiles was done following \citet[][]{2012ApJ...759...98C}. 
In short, broken disks are parametrised in the plane of the galaxy through the generalisation 
of broken exponential functions \citep[][]{2008AJ....135...20E} that are integrated along the line-of-sight. 
Figure~\ref{plot_thin_thick_deprojections} shows the 1D model of $\mu_{3.6\mu \rm m}$ for UGC~7321, 
as well as the deprojected SB radial profiles for the thin and thick disks. 
For comparison, the insets display the same profiles for NGC~4565. 

After deprojection, the multi-band truncation in UGC~7321 was identified 
at $\mu_{3.6 \mu \rm m}$(AB)$\approx 26$ mag arcsec$^{-2}$, which is close to the SB limit of the S$^4$G. 
This means that the survey is not ideally suited for the analysis of truncations in face-on galaxies, 
but sufficiently deep for their identification in edge-on galaxies. 
Deprojected 3.6 $\mu$m SB profiles were converted to surface stellar 
densities ($\Sigma_{\star}$) following \citet{2013ApJ...771...59M}:
\begin{equation}
{\rm log_{10}}(\Sigma_{\star}/[\mathcal{M}_{\odot}\rm \,kpc^{-2}])=16.76-0.4 \cdot  \mu_{3.6\mu \rm m}/[\rm mag\,arcsec^{-2}],
\label{munoz2}
\end{equation}
adopting a stellar mass-to-light ratio $\mathcal{M}_{\star}/L=\Upsilon_{3.6 \rm \mu m}=0.53$ \citep[][]{2012AJ....143..139E}. 
A $30\%$ uncertainty on $\mathcal{M}_{\star}/L$ was assumed \citep[see][]{2012ApJ...744...17M}.

The deprojected stellar density of UGC~7321 at the truncation is $1.5 \pm 0.5 \, \mathcal{M}_{\odot} \, {\rm pc}^{-2}$ 
(grey lines in Fig.~\ref{plot_thin_thick_deprojections}). 
In NGC~4565, the truncation occurs at $3.9 \pm 1.2 \, \mathcal{M}_{\odot} \, {\rm pc}^{-2}$, 
which is slightly larger than the $\Sigma_{\star}$ ($\sim 1 \, \mathcal{M}_{\odot} \, {\rm pc}^{-2}$) implied 
by \citet[][]{2019MNRAS.483..664M} and \citet[][]{2020MNRAS.493...87T}. 
This discrepancy, and the relevance of the local value of $\Sigma_{\star}$ at the truncation, 
are discussed in Sect.~\ref{discussion_section}.
\section{Discussion}\label{discussion_section}

\subsection{Galaxy sizes measured from SF thresholds}\label{SF_threholds_truncation}

Many galaxies show SB truncations in their outskirts \citep[e.g.][]{1970ApJ...160..811F, 1979A&AS...38...15V} 
at optical and NIR wavelengths. These have either been linked to the maximum angular momentum of the protogalactic 
cloud \citep[e.g.][]{1987A&A...173...59V,2012MNRAS.427.1102M} and the presence of disk warps \citep[][]{1987A&A...173...59V} 
or to the presence of a SF threshold \citep[][]{1989ApJ...344..685K}. 
Following the work of \citet[][]{2001ApJ...555..301M}, \citet[][]{2020MNRAS.493...87T} proposed exploring truncations in H$\alpha$ and UV to further constrain the link between galaxy edges and SF thresholds. 
This is tested here using deep UV, optical, and NIR imaging.

The isomass contour at $1 \, \mathcal{M}_{\odot} \, {\rm pc}^{-2}$ has been proposed as a proxy of the 
galaxy edge linked to a SF threshold \citep[][]{2020A&A...633L...3C,2020MNRAS.493...87T} (Sect.~\ref{introduction}). 
We revisited this in two widely studied nearby edge-on galaxies, a MW-type and a low-mass diffuse galaxy, 
with a factor of $2.3$ difference in $V_{\rm c}$. 
Their disk truncations are found, in face-on view, at surface stellar densities of $3.9 \pm 1.2$ and 
$1.5 \pm 0.5 \, \mathcal{M}_{\odot} \, {\rm pc}^{-2}$, respectively. 
\citet[][]{2020MNRAS.493...87T} speculated that the gas density threshold for SF is lower in dwarf galaxies than 
in their more massive counterparts, possibly due to a lower SFE in 
low-mass galaxies \citep[][]{2008AJ....136.2782L}. This is in agreement with our observations. 

We also report the existence of a NUV-[3.6] and NUV-[$grz$] colour upturn beyond the truncation in UGC~7321, 
in both the thin and thick disks (Sect.~\ref{7321_truncation}). 
A similar reddening was found in the NUV-[$gri$] colour profile of the thin disk of NGC~4565 \citep[][]{2019MNRAS.483..664M}. 
These $U-$shaped profiles could be related to trends reported in previous works 
for low-inclination galaxies using optical colours \citep[e.g.][]{2008ApJ...683L.103B,2008ApJ...679L..69A}. 
They may indicate that the truncation is linked to a drop in SFE, but accurately probing the cold gas 
and properly correcting for dust would be needed for a decisive interpretation. 
In addition, \citet[][]{2003AJ....125.2455U} argue that the H{\sc\,i} gas surface density in UGC~7321 is 
systematically below that required for efficient SF based on the dynamical criterion of \citet[][]{1989ApJ...344..685K}. 
Deriving deprojected SFE profiles for edge-on galaxies is highly non-trivial. 
This effort might demand integral field unit data, which would also allow for the recovery of SF histories, 
which is beyond the scope of this paper.

We conclude that $R_{1}$ is an accurate proxy of the disk size of UGC~7321. 
However, notable uncertainties are inherent to the deprojection of $\Sigma_{\star}$ and hence affect its local value at the truncation. 
The calibration of $\mathcal{M}_{\star}/L$ is non-trivial and 
can vary for galaxies of different masses and metallicities \citep[e.g.][]{2018ApJ...865..154H}. 
Also, the scale height may vary with radius (see further discussion in Sect.~\ref{thick_disk_internal}). 
On the other hand, the radius of UGC~7321 is not necessarily representative of other galaxies of similar mass: 
ultra-thin galaxies have been claimed to have larger scale-lengths than ordinary disk galaxies 
given their higher stellar specific angular momentum \citep[][]{2019MNRAS.488..547J}. 
While $R_{1}$ can be used as a disk size definition in massive and faint galaxies, 
it is likely that the value of $\Sigma_{\star}$ at the disk edge is not constant for all masses.

\subsection{Thick disks can form through internal processes only}\label{thick_disk_internal}

Whether thick disks form through internal or external, slow or fast, processes is still debated (Sect.~\ref{introduction}). 
The analysis of UGC~7321 allows us to focus on internal slow processes alone, as this galaxy is isolated, bulgeless, 
and shows no clear signs of environmental activity at the explored SB levels (Sect.~\ref{introduction} and references therein). 
The relevant internal processes proposed include in situ thick disk formation, 
concomitant with the buildup of the galaxy, or secular heating and the radial migration of stars.

Thick disks truncate less frequently than thin disks \citep[][]{2011ApJ...741...28C}. 
When they do, the truncation radius is comparable to that in the thin disk. 
\citet[][]{2007ApJ...667L..49D} reported that the truncation in the low-mass edge-on galaxy NGC~4244 occurs at the same radius for 
young, intermediate age, and old stars, at different heights above the mid-plane (up to 1.5 kpc). 
Based on the analysis of resolved stellar populations, 
\citet[][]{2012ApJ...753..138R} also concluded that the thin and the thick disk truncate at the same radius 
in the face-on spiral galaxy NGC~7793. These observational analyses led to the conclusion that dynamical processes are likely 
responsible for the occurrence of truncations in both thin and thick disks, 
while SF thresholds were discarded as an explanation for this phenomenon \citep[e.g.][]{2007ApJ...667L..49D}. 
If this were indeed the case, the different formation epochs of the two components should place their truncations at different radii.

Interestingly, \citet[][]{2019MNRAS.483..664M} find truncations at and above the mid-plane 
(up to $3 \, {\rm kpc}$) at the same radius in NUV, $gri$, and 3.6 $\mu$m images in two MW-like galaxies; 
these truncations are compatible with a SF threshold. 
NGC$\,$4565's multi-$\lambda$ truncation is, however, not found in SB axial profiles 
averaged over the thick disk (Sect.~\ref{7321_truncation}). 
This is expected, as massive galaxies likely formed their thick disk $\sim 8-10$ Gyr ago \citep[][]{2021A&A...645L..13C}. 
Signatures of a SF threshold in the thick disk could have been washed out due to internal (migration) or 
environmental transformations \citep[NGC$\,$4565 has an asymmetric warp and is interacting;][]{1979A&AS...38...15V}. 
Whether these arguments can be extended to low-mass galaxies is unclear. 
Galaxies with different masses may form thick disks through different paths \citep[][]{2012ApJ...759...98C}.

We have reported the discovery of a truncation linked to a SF threshold in the disk of an edge-on diffuse galaxy. 
The sharp SB drop-off is found at the same location in NIR, optical, NUV, and FUV bandpasses (Sect.~\ref{7321_truncation}). 
Moreover, it is detected at the same radius in both the thin and thick disks, the latter being defined 
from decomposition models of 3.6~$\mu$m images by \citet[][]{2018A&A...610A...5C}. 
This challenges thick disk formation models and gives rise to various interpretations. 
\citet[][]{2021A&A...645L..13C} predict that the age of the youngest stars in thick discs is $\sim 4-6$~Gyr 
for galaxies with a total stellar mass of $\mathcal{M}_{\star}\approx 10^{9} \, \mathcal{M}_{\odot}$. 
Thus, low-mass galaxies such as UGC~7321 may host relatively young thick disks whose SF threshold is 
preserved and similar to that of the thin disk. 
In this sense, our observations are likely a consequence of the pristine nature of UGC~7321 in terms of dynamics and SF, 
relative to giant spirals \citep[][]{1999AJ....118.2751M}. The low level of UV radiation 
detected in the region of the thick disk is likely associated with the emission from H{\sc\,ii} regions 
in the mid-plane of this super-thin galaxy. 
In fact, some UV clumps in the thin disk can have a full-width-at-half-maximum as large as $10-15\arcsec$.

The truncation in a thick disk can also result from both heating and radial and vertical migrations, 
despite the galaxy being isolated \citep[][]{2013MNRAS.433..976R}. 
Thick disks can be made through embedded flares of mono-age stellar populations \citep[][]{2015ApJ...804L...9M}. 
Moreover, simulations by \citet[][]{2012A&A...548A.127M} showed that secular evolution (no external perturbations) 
can form flared disks due to angular momentum redistribution caused by spirals or bars. 
This is consistent with the report of a flared disk in UGC~7321 by \citet[][]{2019A&A...628A..58S}, 
where the used data probe the disk up to $x=2$ arcmin, while the flaring is predicted beyond this point, 
far out in the outskirts (the factor of $\sim 2$ difference in their adopted distance with respect to our work is noted).

If such a flare exists, it should happen at a very low SB ($\mu_{3.6 \mu \rm m}$(AB) > 26 mag arcsec$^{-2}$), 
and its contribution to the thick disk fit would not be large. It is not noticeable from the images presented in Sect.~\ref{data}. 
The fits of the outermost vertical cuts of 3.6~$\mu$m SB performed by \citet[][]{2018A&A...610A...5C} 
indeed yielded $\sim 20\%$ larger scale-heights, for both thin and thick disks, than those of the innermost cuts\footnote{
Inner and outer vertical cuts refer to those averaged over axial ranges 
$0.2 R_{25} < x < 0.5 R_{25}$ and $0.5 R_{25} < x < 0.8 R_{25}$, respectively, where 
$R_{25}$ is the isophotal 25 mag arcsec$^{-2}$ radius in the $B$ band \citep[for further details, see][]{2012ApJ...759...98C}. 
For UGC~7321, inner and outer axial ranges correspond to roughly [3-7.5] kpc and [7.5-12.5] kpc, respectively.}. 
That is, the scale heights moderately increase with increasing axial distance. 
(Unfortunately, these outer fits did not fulfil the Comer\'on et al. quality criteria in the decompositions, 
but the thick disk component is also present in them.) 
We conclude that the outer parts of its thick disk may be biased by the light of a flared thin disk, 
but this effect should be negligible as the thick disk already dominates at 
3.6 $\mu$m SB levels as high as $\mu_{3.6}$(AB) $\approx 24$ mag arcsec$^{-2}$ \citep[][]{2018A&A...610A...5C}.

Bars do play an important role in the redistribution of material throughout the 
disk of massive galaxies \citep[e.g.][]{2006ApJ...645..209D,2016A&A...596A..84D}; this is the case of NGC~4565, 
which hosts a prominent peanut-shaped bulge \citep[][]{1986AJ.....91...65J}. 
Bars are also more frequent than previously thought in low-mass galaxies \citep[][]{2016A&A...587A.160D}. 
\citet[][]{2003A&A...409..485P} provided evidence for peanut-shaped outer isophotes in UGC~7321 from an $R-$band image. 
A bar in UGC~7321, as also inferred from the analysis of its H{\sc\,i} position-velocity diagram \citep[][]{2003AJ....125.2455U}, 
may also be responsible for its stellar migration \citep[but see][]{2014MNRAS.439..929G}. 
Thus, UGC~7321's thick disk may be linked to bar-induced internal dynamics. 
Likewise, the outer reddening of UGC~7321 (colour $U-$shape; Fig.~\ref{plot_thin_thick_S4G}) 
is likely associated with SF thresholds followed by bar-driven outer migrations of stars, as in NGC~4565 \citep[][]{2019MNRAS.483..664M}.

\section{Conclusions}\label{summarysection}

We have reported the discovery of an outer truncation in the SB profile of the diffuse ultra-thin edge-on galaxy UGC~7321 
that is seen in UV ({GALEX} FUV and NUV), optical (DESI \emph{grz}), and NIR (\emph{Spitzer} 3.6 $\mu$m) images. 
The truncation,  detected at the same radius in both the thin and thick disks, 
hints at similar or interconnected (migration) formation mechanisms for both components. 
The truncation occurs at a deprojected stellar surface density of $1.5 \pm 0.5 \, \mathcal{M}_{\odot} \, {\rm pc}^{-2}$, 
in agreement with the theoretical gas density thresholds for SF.
The redder colours beyond the truncation are indicative of the radial migration of stars to the galaxy's outskirts. 
As UGC~7321 is isolated and has no strong signs of accretion, 
our findings are consistent with its thick disk and truncations being formed via internal mechanisms alone.

\begin{acknowledgements}

We thank the anonymous referee for a constructive and detailed report. 
This project has received funding from the European Union’s Horizon 2020 research and innovation programme 
under the Marie Sk$\l$odowska-Curie grant agreement No 893673, 
from the State Research Agency (AEI-MCINN) of the Spanish Ministry of Science and Innovation 
under the grant ``The structure and evolution of galaxies and their central regions'' 
with reference PID2019-105602GB-I00/10.13039/501100011033, 
and under the grant ``Thick discs, relics of the infancy of galaxies" with reference PID2020-113213GA-I00, 
and from IAC project P/300724, financed by the Ministry of Science and Innovation, through the State Budget and by the 
Canary Islands Department of Economy, Knowledge and Employment, through the Regional Budget of the Autonomous Community. 
S.C. is especially grateful to the Natural Sciences and Engineering Research Council of Canada, the Ontario Government, 
and Queen's University for support through various scholarships and grants. 
A.W. acknowledges support from the STFC [ST/S00615X/1]. 
J.H.K. acknowledges support from the ACIISI, Consejer\'{i}a de Econom\'{i}a, 
Conocimiento y Empleo del Gobierno de Canarias and the 
European Regional Development Fund (ERDF) under grant with reference PROID2021010044. 
J.R. acknowledges funding from University of La Laguna through the Margarita Salas Program from the Spanish Ministry of Universities ref. 
UNI/551/2021-May 26, and under the EU Next Generation. 
This research makes use of python (\href{http://www.python.org}{http://www.python.org}) 
and IDL (\href{https://www.harrisgeospatial.com/docs/using_idl_home.html}{https://www.harrisgeospatial.com/docs/using$\_$idl$\_$home.html}).

\end{acknowledgements}

\bibliographystyle{aa}
\bibliography{bibliography}

\end{document}